\documentclass[conference]{IEEEtran}
\ifCLASSINFOpdf
\else
\fi
\usepackage{color}
\usepackage{enumerate}
\usepackage[cmex10]{amsmath} 
\usepackage{siunitx}
\usepackage{import} 
\usepackage{enumitem}
\usepackage{indentfirst} 
\usepackage{standalone}
\usepackage[section]{placeins} 
\usepackage{pdflscape}
\usepackage{tabularx}
\usepackage{algorithm} 
\usepackage{amssymb}
\usepackage{microtype}
\usepackage{soul} 
\usepackage{mathtools}
\usepackage{multirow}
\usepackage{verbatim}
\usepackage{graphicx}
\usepackage{pgfplots}

\usepackage{subfloat} 
\usepackage{verbatim}
\usepackage[utf8]{inputenc}
\usepackage[T1]{fontenc}
\usepackage{amsmath} 
\usepackage[shellescape,latex]{gmp} 
\usepackage[space]{grffile}
\usepackage{booktabs} 
\usepackage{xfrac}
\usepackage{tabularx}
\usepackage{cite}
\usepackage{adjustbox}
\usepackage{verbatim}
\usepackage{url}
\usetikzlibrary{shapes}
\usepackage{multicol, blindtext}    
\usepackage[]{color}
\usepackage{framed}
\usepackage{alltt}
\usepackage[T1]{fontenc}
\usepackage{array}
\usepackage{colortbl}
\usepackage{booktabs}
\usepackage{multirow}
\usepackage{eurosym}
\usepackage{listings}
\usepackage{ifthen}
\usepackage{pgfplots}
\usepackage{pgfplotstable}
\usepackage{pgfcalendar}
\usepackage{pgfgantt}
\usepackage{tikz}
\usetikzlibrary{shapes.geometric, arrows}
\usepackage{longtable}
\usepackage{amsfonts}
\usepackage{subfigure}
\usepackage{float}
\usepackage{mathrsfs}
\usepackage{algpseudocode} 
\usepackage{textcomp}
\usepackage{xspace}
\makeatletter
\newcommand\fs@betterruled{
  \def\@fs@pre{\vspace*{5pt}\hrule height.8pt depth0pt \kern2pt}%
  \def\@fs@post{\kern2pt\hrule\relax}%
  \def\@fs@mid{\kern2pt\hrule\kern2pt}%
  \let\@fs@iftopcapt\iftrue}
\floatstyle{betterruled}
\restylefloat{algorithm}
\makeatother
\pgfplotsset{compat=1.16}
\usepackage{fmtcount}

\usepackage[
  top=0.72in,
  bottom=1in,
  left=0.64in,
  right=0.64in
]{geometry}
\begin{document}

\title{Deadline-Driven Hierarchical Agentic \\ Resource Sharing for AI Services and RAN Functions in AI-RAN
}

\author{
\IEEEauthorblockN{Haiyuan Li, Yulei Wu, Dimitra~Simeonidou}
\IEEEauthorblockA{\textit{Smart Internet Lab, Faculty of Science and Engineering, University of Bristol, BS8 1UB, U.K}\\
E-mail: \{ocean.h.li, y.l.wu, dimitra.simeonidou\}@bristol.ac.uk}
}

\maketitle

\begin{abstract}
AI-RAN consolidates AI services and Radio Access Network (RAN) functions onto a unified, GPU-accelerated infrastructure at the network edge. 
However, compute sharing between real-time RAN functions and highly heterogeneous AI services requires coordination of scheduling decisions at mismatched timescales, and placement adaptation may require service migration across nodes with non-negligible interruptions.
This paper proposes a hierarchical agentic framework (HAF) for compute sharing in AI-RAN that combines a large language model (LLM)-based agent for slow-timescale placement of AI services and RAN functions with a closed-form, deadline-aware convex algorithm for fast-timescale GPU/CPU allocation.
The LLM agent is further equipped with a predictive critic that filters out migrations when the induced service interruption outweighs the expected service-level objective (SLO) benefit.
Experimental results show that HAF reaches 90.0\% overall SLO fulfillment, a 20.5\% improvement over the strongest baseline, and raises AI service request fulfillment from 51\% to 85.3\%.
Further evaluations show that HAF retains its advantage under diverse load conditions, while the critic consistently improves SLO fulfillment across multiple open-source LLM agents.
\end{abstract}


\vspace{-0.1cm}
\section{Introduction} \label{sec:Introduction}
\vspace{-0.1cm}
The rapid proliferation of AI services is pushing inference workloads from centralized cloud data centers toward the network edge, where service quality increasingly depends on proximity to data generation and end users~\cite{lin2025pushing}. In parallel, the radio access network (RAN) itself is becoming AI-native, with machine learning embedded into network management processes~\cite {li2025toward, basaran2025next}. These trends converge at the radio edge, where industry initiatives such as the AI-RAN Alliance and recent SoftBank/NVIDIA field trials have introduced the AI-RAN paradigm, i.e., consolidating AI services and RAN functions onto a unified, GPU-accelerated infrastructure as a cornerstone of future 6G systems~\cite{kundu2025ai}.

Despite its promise, realizing AI-RAN creates direct compute contention rooted in the RAN user-plane data path. The distributed unit (DU) handles the GPU-heavy PHY/MAC processing of end users, and the centralized unit user-plane (CU-UP) processes the CPU-heavy Packet Data Convergence Protocol (PDCP) and forwarding stages, placing both alongside AI services on the same edge networks~\cite{9569133, 10621380}. 
Since AI service requests traverse both the RAN processing chain and the invoked AI service, their end-to-end latency is jointly determined by RAN functions and AI services in the shared compute pool. This coupling requires AI and RAN workloads to be provisioned against a common service-level objective (SLO) budget.
The two classes of services also differ in real-time and scale characteristics. RAN functions are hard real-time, with demand bounded by physical-layer parameters. AI services exhibit more elastic request patterns and vary widely in model size and loading time. Moreover, dynamic RAN traffic and AI demand reshape network resource pressure, requiring adaptive runtime service placement and resource allocation~\cite{feng2026ai}.

Existing compute-management studies address AI-RAN resource contention from three main directions: 
(i) \textit{static partitioning}~\cite{polese2025beyond,kundu2025ai}, such as Multi-Instance GPU (MIG) slices or fixed functional placements, which is simple but may waste capacity under bursty demand; 
(ii) \textit{optimization-based management}, including MILP and heuristic RAN placement~\cite{d2023orchestran} and collaborative AI inference resource optimization~\cite{li2023joint}, yet the coupled contention between AI services and RAN functions over shared GPU, CPU, and memory resources remains insufficiently addressed;
and (iii) \textit{learning-based dynamic scheduling}~\cite{shah2025interplay,shah2025proactive}, which dynamically arbitrates resources but often couples millisecond-scale resource allocation with second-scale service migration in a single loop. 
Moreover, existing studies often treat AI services as homogeneous despite substantial differences in memory footprint and latency targets across large-model and lightweight services~\cite{sun2025review}. They also rarely account for the service interruption caused by migration during placement adaptation, including AI-service model reloading~\cite{fu2024serverlessllm} and RAN-function reinitialization such as DU PHY/MAC stack initialization or CU-UP control-plane re-attachment.

To address these challenges, this paper introduces a hierarchical agentic framework (HAF) for asynchronous and deadline-driven compute sharing between AI services and RAN functions in AI-RAN. The main contributions are summarized as follows:
\begin{itemize}[leftmargin=*]
\vspace{-0.04cm}
    \item \textit{Deadline-driven orchestration:} We formulate AI-RAN compute sharing as a deadline-driven orchestration problem over a shared compute pool, where AI service and background RAN-only requests impose distinct latency budgets.
    The objective maximizes AI service request fulfillment under hard real-time constraints for RAN functions, while accounting for migration-induced deadline violations. 
    \item \textit{Hierarchical agentic framework:} Building on this formulation, we design a two-layer HAF that separates slow-timescale placement and migration from fast-timescale GPU/CPU allocation. A large language model (LLM)-based agent reasons over AI service and RAN function placement, while a closed-form deadline-aware convex allocator reacts to event-level demand. A predictive critic is further incorporated into the slow-timescale placement layer to forecast migration impact and filter suboptimal migrations. 
    \item \textit{Extensive experiments:} Experiments show that HAF achieves 90.0\% overall SLO fulfillment, achieving more than a 20\% improvement over the strongest baseline, and raises AI service request fulfillment from 51\% to 85.3\%. HAF also maintains its advantage across various load levels, while the critic consistently improves SLO fulfillment for different open-source LLM agents.
\end{itemize}

\vspace{-0.1cm}
\section{System Model and Problem Formulation}
\label{sec:system}
\vspace{-0.1cm}
We consider an AI-RAN edge cluster where heterogeneous nodes $\mathcal{N}$ share GPU compute, GPU memory, and CPU resources between AI services and RAN functions. Each node $n$ has GPU capacity $G_n$, GPU memory $V_n$, and CPU capacity $C_n$, and all nodes are interconnected by a fabric with one-way per-hop delay $\delta$. The hosted instance set $\mathcal{S}$ comprises four categories: GPU-bound DU functions $\mathcal{S}^D$ for physical-layer baseband, CPU-bound CU-UP functions $\mathcal{S}^U$ for PDCP and user-plane forwarding, large-AI services $\mathcal{S}^L$ such as LLMs with multi-gigabyte weights and second-scale reload times, and small-AI services $\mathcal{S}^S$ such as vision or embedding models with sub-gigabyte weights and sub-second reload times. Each instance $s$ has a persistent GPU memory footprint $M_s$ for AI model weights or DU PHY/MAC libraries, while CU-UP functions run on the CPU and have $M_s=0$.

Requests are divided into AI service requests $\mathcal{Q}^e$, which traverse RAN functions and an AI service, and RAN-only requests $\mathcal{Q}^r$, which traverse only DU and CU-UP. Each request $q$ arrives at $a_q$, has deadline $\tau_q$, targets instance $s(q)$, and carries per-instance GPU/CPU work $(\Phi_{q,s}^g,\Phi_{q,s}^c)$. Active AI service requests additionally occupy transient KV-cache memory $\gamma_q$ that scales with its output length.
For each AI service request, the DU and CU-UP instances are determined by the user's serving cell, with each cell associated with a DU/CU-UP function pair. AI-service requests are routed to the resident node with the smallest backlog. Let $\mathcal{A}_{n,s}(t)$ denote the active requests served by instance $s$ on node $n$.

\begin{samepage}
For a RAN-only request $q \in \mathcal{Q}^r$, the processing time on RAN function $X \in \{D, U\}$ is denoted $\alpha_q^X$, summing the GPU and CPU contributions on the node $n^X(q) \in \mathcal{N}$ hosting $s^X(q)$. We write $y_{n,s}(t) \in \{0,1\}$ for the residence indicator, with $y_{n,s}(t) = 1$ when instance $s$ is resident on node $n$ at time $t$. The GPU and CPU capacities allocated to $s$ on $n$ at time $t$ are $g_{n,s}(t)$ (FLOPs/s) and $c_{n,s}(t)$ (cores), with $g_{n,s}(t) = c_{n,s}(t) = 0$ whenever $y_{n,s}(t) = 0$. Accordingly,
\vspace{-0.15cm}
\begin{equation}
    \alpha_q^X = \frac{\Phi_{q, s^X(q)}^g}{g_{n^X(q), s^X(q)}(t)} + \frac{\Phi_{q, s^X(q)}^c}{c_{n^X(q), s^X(q)}(t)},
\end{equation}
\vspace{-0.15cm}
For an AI service request $q \in \mathcal{Q}^e$, the processing time on the invoked AI service is $\beta_q$, defined analogously on $(n_q, s(q))$. 
\end{samepage}
The end-to-end latency is then
\begin{equation}
    T_q = \begin{cases}
    \alpha_q^D + \alpha_q^U + \delta_q, & q \in \mathcal{Q}^r, \\
    \delta_q + \beta_q, & q \in \mathcal{Q}^e,
    \end{cases}
    \label{eq:e2e_latency}
\end{equation}
where $\delta_q$ aggregates intra-cluster transport delay, with each inter-node hop contributing $\delta$. For AI service requests, the RAN-stage packet-processing delay is included in $\delta_q$, while AI-service compute contention is captured through $\beta_q$ and the shared node-level allocation. Multiple requests assigned to the same $(n,s)$ share the allocated GPU/CPU capacity, and their aggregate residual work forms the service backlog.

In addition to per-request service dynamics, instance residence can change through migration actions. When an action changes $y_{n,s}:0\to1$, instance $s$ undergoes a reconfiguration interval $R_s$ at the destination node, during which it is unavailable and requests routed to it are delayed.
The value of $R_s$ depends on the migrated instance type. Large-model AI services typically require second-scale model reloading, lightweight AI services have shorter loading delays, and RAN functions involve reinitialization steps such as DU PHY/MAC stack initialization or CU-UP control-plane re-attachment. This waiting time contributes to the realized service latency and therefore affects deadline fulfillment in Eq.~\eqref{eq:e2e_latency}.

At every time $t$, the GPU and CPU allocations on each node must remain within the node's capacity.
\begin{align}
    \sum_{s \in \mathcal{S}} g_{n,s}(t) \le G_n,
    \quad
    \sum_{s \in \mathcal{S}} c_{n,s}(t) \le C_n, \quad \forall n, t.
    \label{eq:gpu_cpu_cap}
\end{align}
Additionally, the GPU memory budget on each node covers both resident model weights and the transient KV cache held by the requests $q \in \mathcal{A}_{n,s}(t)$ currently being served on $(n,s)$.
\begin{equation}
    \sum_{s \in \mathcal{S}} M_s y_{n,s}(t) + \sum_{s \in \mathcal{S}} \sum_{q \in \mathcal{A}_{n,s}(t)} \gamma_q \le V_n, \quad \forall n, t.
    \label{eq:mem_cap}
\end{equation}

Given the latency model and resource constraints above, we formulate a deadline-driven orchestration problem over instance placement and per-node GPU/CPU allocation. The formulation follows the service priority in AI-RAN deployments, where RAN functions deliver foundational connectivity with hard real-time requirements, while AI services are elastic tenants whose deadline fulfillment is maximized subject to the RAN guarantee. Accordingly, we treat RAN-level deadline satisfaction as a constraint and end-to-end fulfillment of AI service requests as the objective. The decision variables span the placement of all instances in $\mathcal{S}$ and the per-instance GPU/CPU allocations on each node. The problem is formalized as
\begin{subequations}\label{eq:problem}
\begin{align}
    \max_{\{y_{n,s}(t),\, g_{n,s}(t),\, c_{n,s}(t)\}}
    \quad & \frac{1}{|\mathcal{Q}^e|}
    \sum_{q \in \mathcal{Q}^e}
    \mathbf{1}\bigl[T_q \le \tau_q\bigr]
    \label{eq:obj}\\[3pt]
    \text{s.t.}\quad
    & T_q \le \tau_q, \quad \forall q \in \mathcal{Q}^r,
    \label{eq:ran_hard}\\
    & \eqref{eq:gpu_cpu_cap}, \eqref{eq:mem_cap}.
\end{align}
\end{subequations}
Objective~\eqref{eq:obj} maximizes AI-service request deadline fulfillment, while constraint~\eqref{eq:ran_hard} enforces hard RAN deadlines and is implemented through capacity floors in Section~\ref{sec:method-C}. 
The decisions in problem~\eqref{eq:problem} are coupled across time by migration delays and across variables because placement and allocation jointly determine fulfillment. At each epoch, the placement space already scales as $2^{|\mathcal{N}|\cdot|\mathcal{S}|}$, and this search further compounds over the planning horizon. Because the expected fulfillment under each placement has no closed-form expression, it is intractable to solve the problem as a monolithic optimization.

\vspace{-0.1cm}
\section{Hierarchical Agentic Framework}
\label{sec:method}

\begin{figure}[t]
    \centering
    \includegraphics[width=0.98\linewidth]{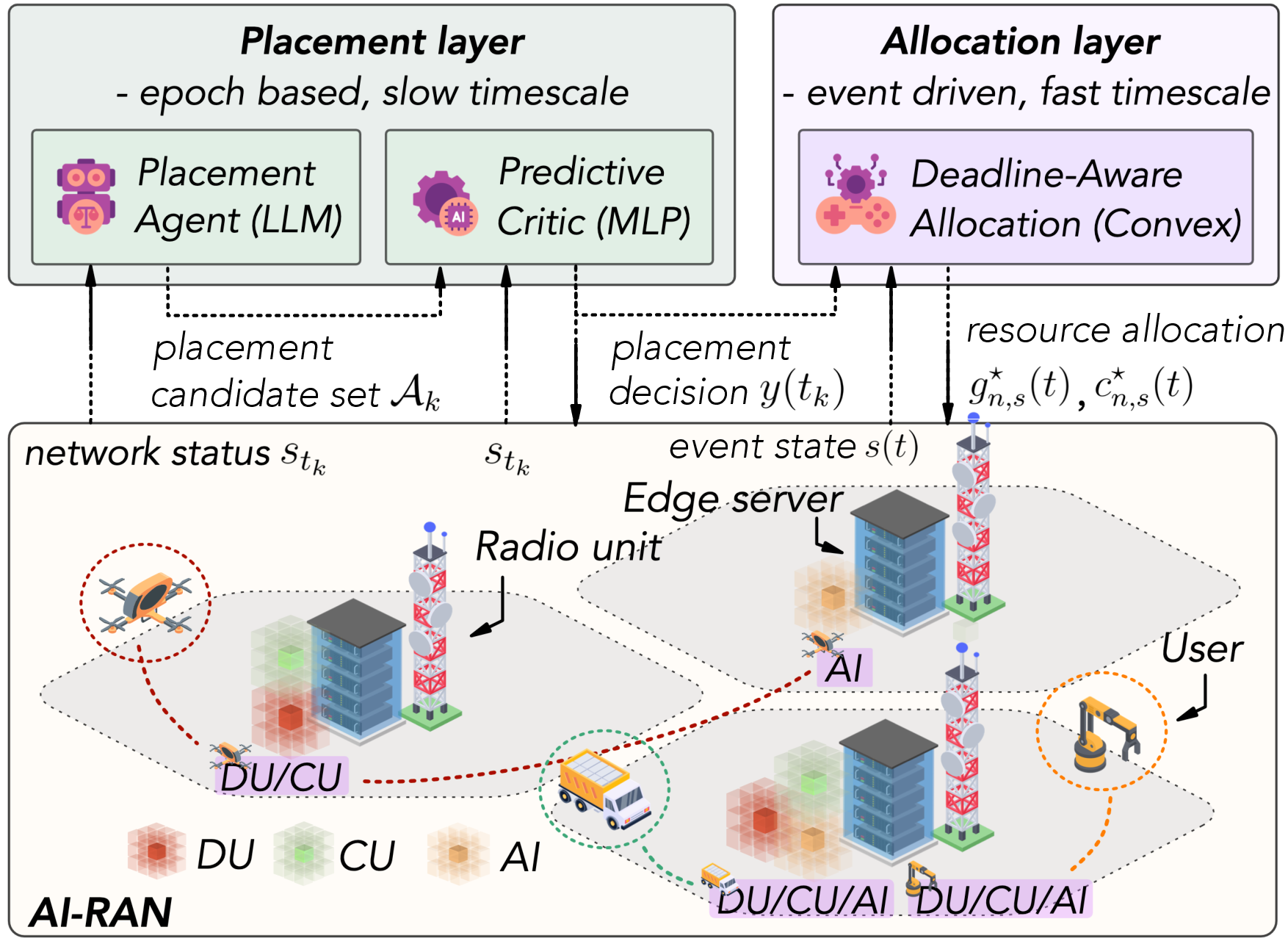}
    \vspace{-0.2cm}
    \caption{HAF architecture: a slow, epoch-based placement layer and a fast, event-driven allocation layer over AI-RAN.}
    \vspace{-0.6cm}
    \label{fig:solution}
\end{figure}

The intractability of the problem originates from two structural features. Placement is held over seconds-long intervals, while resource allocation must react to bursty request arrivals within milliseconds.
Building on this separation, we propose a hierarchical agentic framework (HAF) that organizes control into two layers as shown in Fig.~\ref{fig:solution}. The \emph{placement layer} operates at discrete epochs $\{t_k\}$ separated by a fixed interval $\Delta$ and handles the integer placement variables $\{y_{n,s}(t_k)\}$ over the interval $[t_k, t_{k+1})$. It consists of an LLM-based agent that proposes a set of candidate migration actions, paired with a critic that scores the resulting placements through a learned surrogate of the objective.
The \emph{allocation layer} operates on an event-driven basis within each interval, takes the committed placement as input, and computes $\{g_{n,s}(t), c_{n,s}(t)\}$ through a closed-form active-set allocation under per-node GPU/CPU capacity constraints.

\vspace{-0.2cm}
\subsection{Placement Layer: Agentic Candidate Generation}
At each epoch, the placement layer considers a reduced form of problem~\eqref{eq:problem}, where each candidate placement is evaluated under the allocation-layer response (Section~\ref{sec:method-C}). The reduced placement problem is
\begin{equation}
    y(t_k) = \arg\max_{y \in \mathcal{Y}(y(t_k^-))} V(y, s_{t_k}),
    \label{eq:slow_problem}
\end{equation}
where $s_{t_k}$ is the system state at $t_k$, $\mathcal{Y}(y(t_k^-))$ is the set of placements reachable from the inherited configuration under constraints~\eqref{eq:gpu_cpu_cap}--\eqref{eq:mem_cap}, and
\begin{equation}
\begin{aligned}
V(y,s_{t_k}) ={}& \mathbb{E}\Biggl[
\frac{1}{\left|\mathcal{Q}^e_{[t_k,t_{k+1})}\right|}
\sum_{q \in \mathcal{Q}^e_{[t_k,t_{k+1})}}
\mathbf{1}[T_q \le \tau_q]
\\ &\qquad\qquad
\Big|\, y(t_k)=y,\ s_{t_k}
\Biggr]
\end{aligned}
\label{eq:slow_value}
\end{equation}
is the expected AI-service request deadline-fulfillment rate over the next placement interval under the allocation-layer response.

In response, we decompose the placement-layer decision into candidate action generation, LLM-based shortlisting, and critic-based selection. The placement layer searches over migration actions whose application induces candidate placement configurations. To control the combinatorial growth of $\mathcal{Y}$, we form the candidate action set $\mathcal{M}_k$ at each epoch $t_k$ from feasible single-instance migrations relative to the inherited placement $y(t_k^-)$.
Let $\mathcal{S}^M \subseteq \mathcal{S}$ denote the set of instance categories eligible for migration, and let $\Pi(y,a)$ denote the placement obtained by applying migration action $a$ to placement $y$. For each $s \in \mathcal{S}^M$ and destination $n' \in \mathcal{N}\setminus\{n(s)\}$, the migration $a=(s,n(s)\to n')$ is included in $\mathcal{M}_k$ if the resulting placement $y'=\Pi(y(t_k^-),a)$ satisfies~\eqref{eq:mem_cap} and $s$ is not already undergoing reconfiguration. Including the no-migration option gives $|\mathcal{M}_k| \le |\mathcal{S}^M|(|\mathcal{N}|-1)+1.$
This single-instance restriction reduces the per-epoch search from $2^{|\mathcal{N}|\cdot|\mathcal{S}|}$ to $O(|\mathcal{S}^M|\cdot|\mathcal{N}|)$, serializes migration-induced offline windows $R_s$, and preserves reachability over the planning horizon through successive commits.

Given this candidate action set, we introduce an LLM-based agent to shortlist the most promising actions. The agent is guided by a structured prompt with three components. The first is a system policy that translates the formulation into ordered decision priorities: protecting $\mathcal{Q}^r$ deadline satisfaction, improving $\mathcal{Q}^e$ end-to-end fulfillment, and accounting for the reconfiguration cost $R_s$. These priorities guide the agent toward migrations that relieve GPU/CPU pressure on RAN-critical nodes, move AI services toward nodes with available GPU, CPU, and VRAM capacity, and avoid migrations whose offline cost is unlikely to be recovered.
The second component is a per-epoch state snapshot covering feasibility and contention, including node utilization, RAN-floor utilization, VRAM headroom, active requests, resident services, migration state, backlog, request memory footprint, and destination-node headroom for each candidate in $\mathcal{M}_k$.
The last is the candidate action set $\mathcal{M}_k$, supplied as a list of migration identifiers the agent is allowed to select from. To improve robustness against single-point ranking errors, the agent returns an ordered shortlist $\mathcal{A}_k \subseteq \mathcal{M}_k$ of up to $K$ candidates,
\begin{equation}
    \mathcal{A}_k = \pi_\text{LLM}\!\left(s_{t_k}, \mathcal{M}_k\right), \quad |\mathcal{A}_k| \le K,
    \label{eq:llm_select}
\end{equation}
where $\pi_\text{LLM}$ denotes the prompted LLM agent. 
The ordered shortlist $\mathcal{A}_k$ narrows the candidate action set for subsequent evaluation.

\vspace{-0.1cm}
\subsection{Placement Layer: Predictive Critic}
\label{sec:method-B}
The critic provides a learned surrogate $\hat{r}_\theta$ for the expected fulfillment value $V$ in Eq.~\eqref{eq:slow_problem}. With parameters $\theta$, it maps each state-action pair to a class-resolved fulfillment forecast over the next placement interval:
\begin{equation}
    \hat{r}_\theta: (s_{t_k}, a_k^{(j)}) \mapsto \bigl(\hat{r}^L, \hat{r}^S, \hat{r}^R\bigr) \in [0,1]^3,
    \label{eq:critic_map}
\end{equation}
where the three components are the predicted deadline fulfillment rates over $[t_k,t_{k+1})$ for large-AI, small-AI, and RAN-only requests, respectively. 

The critic is trained offline by supervised regression on placement-epoch samples $(s_{t_k}, a_k, r_k)$, where $a_k$ is the candidate migration action applied at state $s_{t_k}$ and $r_k$ is the class-resolved fulfillment rate observed over $[t_k,t_{k+1})$ after applying $a_k$.
Since this interval includes the migration delay $R_s$ and its effect on request latency, $r_k$ captures the net outcome of each candidate migration:
\begin{equation}
    \mathcal{L}(\theta) = \mathbb{E}\!\left[\|\hat{r}_\theta(s,a)-r\|_2^2\right].
    \label{eq:critic_loss}
\end{equation}
It is frozen at deployment. 
At each placement-layer decision, writing $\mathcal{A}_k = (a_k^{(1)}, \ldots, a_k^{(|\mathcal{A}_k|)})$, 
the critic scores every candidate in $\mathcal{A}_k$ and selects
\begin{equation}
    j^\star = \arg\max_{j \in \{1, \ldots, |\mathcal{A}_k|\}} 
    \bar{r}\bigl(\hat{r}_\theta(s_{t_k}, a_k^{(j)})\bigr),
    \label{eq:critic_select}
\end{equation}
where $\bar{r}(\cdot)$ is a weighted mean of the three-component forecast, with weights reflecting the relative urgency of each request class. 
The committed placement for the upcoming interval is
\begin{equation}
    y_{n,s}(t_k) = \bigl[\Pi\bigl(y(t_k^-), a_k^{(j^\star)}\bigr)\bigr]_{n,s}, 
    \quad \forall n \in \mathcal{N},\ s \in \mathcal{S}.
    \label{eq:commit_placement}
\end{equation}
Therefore, the critic forecast reflects both the expected SLO benefit of a migration and the service interruption induced by that migration.

\subsection{Allocation Layer: Deadline-Aware GPU/CPU Allocation}
\label{sec:method-C}
Given the placement committed for interval $[t_k,t_{k+1})$, the allocation layer reacts to request arrivals, completions, and deadline updates by adjusting the per-instance GPU and CPU allocations $\{g_{n,s}(t),c_{n,s}(t)\}$. For node $n$, let $\mathcal{S}_n(t)=\{s:y_{n,s}(t)=1\}$ denote the set of resident instances. Because the indicator objective in~\eqref{eq:obj} is non-smooth and unsuitable for event-driven allocation, the allocation layer uses a differentiable deadline-weighted processing-time surrogate. The surrogate prioritizes the instances with larger residual workload and tighter deadlines, while the hard RAN deadline constraint~\eqref{eq:ran_hard} is enforced through minimum capacity floors.

Let $\Phi_{q,s}^{g,\mathrm{rem}}(t)$ and $\Phi_{q,s}^{c,\mathrm{rem}}(t)$ denote the residual GPU and CPU work of request $q$ on instance $s$. The aggregate active workload at $(n,s)$ is
\begin{equation}
    \Psi_{n,s}^g(t) = \sum_{q \in \mathcal{A}_{n,s}(t)} \Phi_{q,s}^{g,\mathrm{rem}}(t),
    \
    \Psi_{n,s}^c(t) = \sum_{q \in \mathcal{A}_{n,s}(t)} \Phi_{q,s}^{c,\mathrm{rem}}(t).
    \label{eq:aggregate_work}
\end{equation}
The urgency of active requests at $(n,s)$ is measured by
\begin{equation}
    \omega_{n,s}(t) =
    \sum_{q \in \mathcal{A}_{n,s}(t)}
    \frac{1}{\max(\tau_q-(t-a_q),\,\epsilon)},
    \label{eq:urgency_weight}
\end{equation}
where $\epsilon>0$ avoids division by zero. Requests closer to their deadlines therefore contribute larger weights.
Together, these workload and urgency terms define each resident instance's priority in the allocation objective.

For each resident RAN function, a capacity floor is assigned on its dominant resource. The floor is determined by the remaining RAN-only workload and the minimum remaining time before the deadline among pending RAN-only requests. Let $\mathcal{Q}_{n,s}^r(t)$ denote the set of pending RAN-only requests assigned to instance $s$ on node $n$. If $\mathcal{Q}_{n,s}^r(t)=\emptyset$, the corresponding floor is zero. Otherwise, for each DU instance $s\in\mathcal{S}^D\cap\mathcal{S}_n(t)$, the GPU floor is
\begin{equation}
    \underline{g}_{n,s}(t) =
    \frac{\Psi_{n,s}^g(t)}
    {\displaystyle\min_{q \in \mathcal{Q}_{n,s}^r(t)}
    \bigl(\tau_q-(t-a_q)-\delta-\hat{\alpha}_q^{\mathrm{down}}\bigr)},
    \label{eq:ran_floor_gpu}
\end{equation}
where $\hat{\alpha}_q^{\mathrm{down}}$ is the estimated processing time of downstream RAN stages that request $q$ has yet to traverse. For DU functions, it corresponds to the expected CU-UP processing time; for CU-UP functions, $\hat{\alpha}_q^{\mathrm{down}}=0$. The CPU floor $\underline{c}_{n,s}(t)$ for each CU-UP instance $s\in\mathcal{S}^U\cap\mathcal{S}_n(t)$ is defined analogously by replacing $\Psi_{n,s}^g(t)$ with $\Psi_{n,s}^c(t)$. If the minimum remaining time in the denominator is non-positive, the current placement is infeasible with respect to the RAN deadline constraint.
With these workload, urgency, and floor terms, the per-node allocation problem is
\begin{subequations}\label{eq:fast_problem}
    \begin{align}
        \min_{\{g_{n,s},\, c_{n,s}\}} \quad &
        \sum_{s \in \mathcal{S}_n(t)}
        \omega_{n,s}(t)
        \left(
        \frac{\Psi_{n,s}^g(t)}{g_{n,s}} + \frac{\Psi_{n,s}^c(t)}{c_{n,s}}
        \right)
        \label{eq:fast_obj} \\
        \text{s.t.}\quad
        & \sum_{s \in \mathcal{S}_n(t)} g_{n,s} \le G_n,\
          \sum_{s \in \mathcal{S}_n(t)} c_{n,s} \le C_n,
          \label{eq:fast_resource_cap} \\
        & g_{n,s} \ge \underline{g}_{n,s}(t), \quad \forall s \in \mathcal{S}^D \cap \mathcal{S}_n(t), \label{eq:fast_floor_g}\\
        & c_{n,s} \ge \underline{c}_{n,s}(t), \quad \forall s \in \mathcal{S}^U \cap \mathcal{S}_n(t).
        \label{eq:fast_floor_c}
    \end{align}
\end{subequations}
Constraints in Eq.~\eqref{eq:fast_resource_cap} correspond to the per-node capacity constraints in~\eqref{eq:gpu_cpu_cap}, while constraints~\eqref{eq:fast_floor_g}--\eqref{eq:fast_floor_c} implement the RAN deadline protection required by~\eqref{eq:ran_hard}.

Problem~\eqref{eq:fast_problem} is convex because each term $\Psi/g$ or $\Psi/c$ is convex over positive allocations and all capacity and floor constraints are linear. Since the objective is additive in the GPU and CPU variables, the allocation problem can be decomposed into independent GPU and CPU sub-problems. For the GPU sub-problem, KKT stationarity yields the square-root workload-urgency proportionality for instances not fixed at their floors:
\begin{equation}
    g_{n,s} \propto \sqrt{\omega_{n,s}(t)\Psi_{n,s}^g(t)}.
    \label{eq:kkt_sqrt_rule}
\end{equation}
The floor constraints are handled by active-set clipping. Let $\mathcal{U}_n^g(t)$ denote the final active set of GPU instances whose square-root allocation is above their floor, and let $\mathcal{B}_n^g(t)$ denote the remaining instances fixed at their GPU floors. The resulting GPU allocation is
\begin{equation}
g_{n,s}^{\star}(t)=
\begin{cases}
\underline{g}_{n,s}(t), & s\in\mathcal{B}_n^g(t),\\[3pt]
\tilde{G}_n^g(t)
\dfrac{
\sqrt{\omega_{n,s}(t)\Psi_{n,s}^g(t)}
}{
\sum_{s'\in\mathcal{U}_n^g(t)}
\sqrt{\omega_{n,s'}(t)\Psi_{n,s'}^g(t)}
}, & s\in\mathcal{U}_n^g(t),
\end{cases}
\label{eq:fast_gpu_solution}
\end{equation}
where $\underline{g}_{n,s}(t)=0$ for non-DU instances and
\begin{equation}
    \tilde{G}_n^g(t)
    =
    G_n-\sum_{s\in\mathcal{B}_n^g(t)}\underline{g}_{n,s}(t)
\end{equation}
is the GPU capacity remaining after floor-bound instances are fixed. If the denominator in~\eqref{eq:fast_gpu_solution} is zero, only the required floor allocations are applied. The CPU allocation follows analogously using $\underline{c}_{n,s}(t)$ and $C_n$. This closed-form active-set allocation serves as the fast-timescale component used in the placement-layer reduction in Eq.~\eqref{eq:slow_problem}.

\section{Experimental Evaluation}
\label{sec:experiments}

\begin{table}[!b]
\vspace{-0.4cm}
\centering
\caption{Simulation parameters.}
\vspace{-0.2cm}
\label{tab:sim_params}
\scriptsize
\setlength{\tabcolsep}{3pt}
\renewcommand{\arraystretch}{1}
\begin{tabular*}{\columnwidth}{@{\extracolsep{\fill}}ll@{}}
\toprule
\multicolumn{2}{@{}l@{}}{\textit{Infrastructure}} \\
\midrule
Compute nodes & 6 (2 GPU-heavy, 2 CPU-heavy, 2 balanced) \\
Transport delay $\delta$ & 200\,$\mu$s \\
\midrule
\multicolumn{2}{@{}l@{}}{\textit{Instances}} \\
\midrule
DU functions & 6 (one per cell) \\
CU-UP functions & 6 (one per cell) \\
Large-AI services & 2 \\
Small-AI services & 4 \\
Large-AI model weight $M_s$ & 28\,GB \\
Small-AI model weight $M_s$ & $<$ 1\,GB \\
Per-request KV cache $\gamma_q$ (large-AI) & 0.4--0.6\,GB \\
Migration delay (large-AI) & $\sim$\,8\,s \\
Migration delay (small-AI) & $\sim$\,0.5\,s \\
Migration delay (RAN functions) & $\sim$\,0.05\,s \\
\midrule
\multicolumn{2}{@{}l@{}}{\textit{Workload}} \\
\midrule
$\mathcal{Q}^e$ source & Azure LLM inference trace \\
$\mathcal{Q}^r$ deadlines & 1\,ms (URLLC), 4\,ms (eMBB) \\
$\mathcal{Q}^e$ deadlines & 100\,ms -- a few seconds \\
\midrule
\multicolumn{2}{@{}l@{}}{\textit{HAF parameters}} \\
\midrule
LLM agents (ablation) & qwen3:32b, qwen2.5:72b, gpt-oss:20b, \\
& gpt-oss:120b, deepseek-r1:70b \\
Placement epoch interval $\Delta$ & 5\,s \\
Allocation timescale & Event-driven \\
Candidates per epoch $K$ & 3 \\
Feasible candidate action set $|\mathcal{M}_k|$ & $\le |\mathcal{S}_{\text{movable}}|\cdot(|\mathcal{N}|-1)$ \\
Critic & 2-layer MLP, frozen at deployment \\
\bottomrule
\end{tabular*}
\end{table}

We build a discrete-event simulation environment with six heterogeneous nodes connected through a full-mesh fabric with one-way transport delay $\delta$. The compute pool hosts DU and CU-UP RAN functions, large-AI services for long-context LLM inference, and small-AI services for lightweight vision and embedding workloads. AI service requests $\mathcal{Q}^e$ are drawn from the Azure LLM inference trace~\cite{stojkovic2025dynamollm}, split chronologically, and mapped to small-AI and large-AI services. Per-request compute and memory follow from prompt and response lengths. Background $\mathcal{Q}^r$ requests are synthetic with URLLC and eMBB hard deadlines per 3GPP TR 38.913~\cite{3gpp_38913}. 
The placement layer is evaluated with multiple open-source LLM agents that generate candidate migration actions at each placement epoch, and a frozen MLP critic scores the resulting candidates. Unless otherwise stated, the main results use the best-performing LLM agent selected from the ablation in Table~\ref{tab:llm_ablation}.

\subsubsection{Ablation study} 
We first evaluate HAF across open-source LLM agents to select the agent used in subsequent experiments and quantify the contribution of critic filtering against \textit{HAF-NoCritic}. The evaluation is performed at $\rho = 1.0$, where $\rho$ denotes the AI demand-to-capacity ratio after RAN floor reservation. Specifically, $\rho = \lambda \bar{W} / G$, where $\lambda$ is the aggregate AI service request arrival rate, $\bar{W}$ is the mean per-request GPU work, and $G$ is the cluster GPU capacity available to AI services after the RAN floor reservation in~\eqref{eq:ran_floor_gpu}. Thus, $\rho \approx 1.0$ represents a busy operating point where AI demand approximately matches the effective AI-serving capacity provisioned by the operator for peak-load periods.
As shown in Table~\ref{tab:llm_ablation}, the critic reduces the total migration count for every LLM agent by rejecting migrations whose reconfiguration delay would outweigh their expected gain. This migration-aware gating improves overall SLO fulfillment across all LLM agents, with gains ranging from 1.0\% to 9.1\%. Among the five evaluated LLM agents, qwen3:32b paired with the critic delivers the best overall performance and is used for the subsequent HAF results.

\begin{table}[t]
\centering
\caption{Critic ablation across open-source LLM agents at $\rho = 1.0$. Mig (L/tot) denotes large-AI/total committed migrations.}
\vspace{0cm}
\label{tab:llm_ablation}
\scriptsize
\setlength{\tabcolsep}{3pt}
\renewcommand{\arraystretch}{1}
\begin{tabular*}{\columnwidth}{@{\extracolsep{\fill}}lccccc@{}}
\toprule
 & \multicolumn{2}{c}{HAF (+Critic)} & \multicolumn{2}{c}{HAF-NoCritic} & Critic \\
\cmidrule(lr){2-3} \cmidrule(lr){4-5}
LLM agent & Overall & Mig (L/tot) & Overall & Mig (L/tot) & gain \\
\midrule
qwen3:32b          & 90.0\% & 1/9  & 84.7\% & 2/16 & $+6.3\%$ \\
gpt-oss:20b        & 89.6\% & 1/6  & 84.2\% & 2/16 & $+6.5\%$ \\
qwen2.5:72b        & 86.2\% & 1/8  & 83.5\% & 2/16 & $+3.2\%$ \\
deepseek-r1:70b    & 82.2\% & 1/14 & 81.4\% & 2/16 & $+1.0\%$ \\
gpt-oss:120b       & 88.2\% & 1/10 & 80.8\% & 2/16 & $+9.1\%$ \\
\bottomrule
\end{tabular*}
\vspace{-0.4cm}
\end{table}

\begin{table}[t]
\centering
\caption{SLO fulfillment and migration count.}
\vspace{-0.2cm}
\label{tab:main_results}
\scriptsize
\setlength{\tabcolsep}{2.5pt}
\renewcommand{\arraystretch}{1}
\begin{tabular*}{\columnwidth}{@{\extracolsep{\fill}}lcccccc@{}}
\toprule
Method & Overall & $\mathcal{Q}^r$ fulfill. & $\mathcal{Q}^e$ fulfill. & Large-AI & Small-AI & Mig. (L/tot) \\
\midrule
HAF-Static          & 74.1\% & 96.6\% & 50.8\% &  0.4\% & 100.0\% &  0/0 \\
Round-Robin     & 74.3\% & 97.1\% & 50.8\% &  0.4\% & 100.0\% &  0/0 \\
Lyapunov        & 74.7\% & 97.1\% & 51.6\% &  1.9\% & 100.0\% &  0/7 \\
Game Theory     & 74.7\% & 97.9\% & 50.8\% &  0.4\% &  99.9\% & 0/16 \\
CAORA\cite{shah2025proactive} & 74.3\% & 97.1\% & 50.8\% &  0.5\% & 100.0\% &  0/0 \\
HAF (ours)      & \textbf{90.0\%} & 94.5\% & \textbf{85.3\%} & \textbf{70.4\%} & 99.9\% & 1/9 \\
\bottomrule
\end{tabular*}
\vspace{-0.5cm}
\end{table}

\subsubsection{Overall framework effectiveness}
To evaluate HAF's effectiveness, we compare HAF with the five baselines on the same workload at $\rho \approx 1.0$. These include \textit{(i)} \textit{HAF-Static:} fixed initial placement paired with HAF's allocation layer. The slow-timescale adaptation is removed;
\textit{(ii)} \textit{Round-Robin:} fixed placement with round-robin dispatch and equal-share residual allocation;
\textit{(iii)} \textit{Lyapunov:} a single-layer reactive baseline using drift-plus-penalty control for placement and MaxWeight allocation over the residual capacity;
\textit{(iv)} \textit{Game Theory:} a single-layer market baseline using best-response placement and proportional market clearing over the residual capacity; and \textit{(v)} \textit{CAORA}~\cite{shah2025proactive}: a DRL baseline reproducing the SAC allocation policy. Designed for a single MIG-partitioned GPU, it outputs one scalar $\alpha \in [0,1]$ splitting compute between RAN and AI. We apply $\alpha$ per-node, letting either class take full capacity where it alone resides. Placement is held static, matching the original allocation-only design.
For fairness, all baselines use the same RAN floor reservations defined around Eq.~\eqref{eq:ran_floor_gpu}, so that constraint~\eqref{eq:ran_hard} is enforced consistently.

As shown in Table~\ref{tab:main_results}, HAF delivers 90.0\% overall SLO fulfillment, a 20.5\% relative improvement over the strongest baseline, while HAF-Static, Round-Robin, Lyapunov, Game Theory, and CAORA cluster between 74.1\% and 74.7\%.
The remaining metrics in Table~\ref{tab:main_results} separate the effects of instance placement and resource allocation. The no-migration baselines, HAF-Static, Round-Robin, and CAORA, reuse HAF's closed-form allocation layer, equal-share residual allocation, and the SAC-learned $\alpha$-split, respectively, yet achieve similar SLO fulfillment rates of 74.1\%, 74.3\%, and 74.3\%. This indicates that allocation sophistication alone cannot compensate for an unfavorable instance placement. The migration-capable baselines provide a complementary comparison. Lyapunov and Game Theory perform 7 and 16 migrations, respectively, yet improve only marginally because their migrations are confined to DU, CU-UP, and small-AI services, and the large-AI placement remains unchanged. By contrast, HAF identifies large-AI as the binding service, updates its placement through the LLM-based slow-timescale agent, and optimizes resource allocation through the fast-timescale convex layer. This two-layer coordination yields substantially higher SLO fulfillment.

\begin{figure}[t]
\centering
\includegraphics[width=0.48\columnwidth]{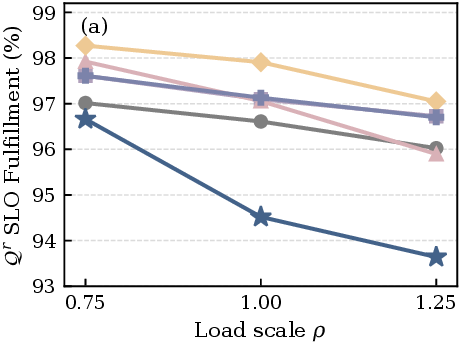}
\includegraphics[width=0.48\columnwidth]{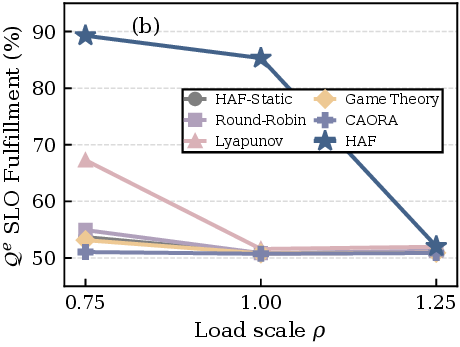}
\vspace{-0.2cm}
\caption{Load sweep across $\rho \in \{0.75, 1.0, 1.25\}$. $\mathcal{Q}^r$ fulfillment stays above 94\% for all methods at all load points, while $\mathcal{Q}^e$ fulfillment separates strongly at $\rho = 0.75$ and $\rho = 1.0$, then converges at $\rho = 1.25$ as the system becomes capacity-limited.}
\label{fig:load_sweep}
\vspace{-0.5cm}
\end{figure}

\subsubsection{Load sensitivity}
To evaluate HAF beyond the peak-hour operating point, we further sweep $\rho$ over $\{0.75,\,1.0,\,1.25\}$. Here, $\rho=0.75$ represents moderate demand with cluster headroom, typical of off-peak office or residential hours, while $\rho=1.25$ represents surge demand beyond peak capacity, such as flash crowds or viral AI request spikes.
At each load point, the AI service and RAN-only request arrival rates are scaled by the same factor, so that $\rho$ reflects the overall network load rather than isolated AI demand. The request count is adjusted to $15,000 / 20,000 / 25,000$ to keep the simulation horizon comparable across load points.
As shown in Fig.~\ref{fig:load_sweep}, HAF attains 89.3\% and 85.3\% $\mathcal{Q}^e$ fulfillment at $\rho = 0.75$ and $\rho = 1.0$, respectively, achieving more than 30\% relative gain over the best baseline. At $\rho = 1.25$, HAF converges with the baseline level at around 52\%. This result indicates a capacity-saturated regime in which aggregate AI demand exceeds the cluster's effective serving capacity and limits the optimization space.

\vspace{-0.2cm}
\section{Conclusion} \label{sec:Conclusion}
This paper presented HAF, a hierarchical agentic framework for deadline-driven compute sharing between AI services and RAN functions in AI-RAN. By separating slow placement from fast GPU/CPU allocation and using a predictive critic to gate costly migrations, HAF achieves 90.0\% overall SLO fulfillment, improves over the strongest baseline by 20.5\%, and raises AI service request fulfillment from approximately 51\% to 85.3\%. The critic further provides a 6.3\% relative SLO improvement over HAF-NoCritic while reducing total migrations by 44\%. Additional evaluations show that HAF maintains its advantage before capacity saturation and that the critic consistently improves SLO fulfillment across different open-source LLM agents.

\section*{Acknowledgments}
\small{This work was supported by the UK Engineering and Physical Sciences Research Council (EPSRC) grant EP/Y037243/1, EP/X04047X/2 for the TITAN Telecoms Hub and the Federated Telecoms Hubs, and grant EP/Y036514/1 for the JOINER project.
}

\bibliographystyle{IEEEtran}
\small{\bibliography{bib}}

\end{document}